\documentstyle[12pt,epsfig]{article}
\begin{document}
\title{\bf{Dissipative Abelian Sandpiles and Random Walks}}
\author{C. Vanderzande, F. Daerden \\
 {\it Departement WNI, Limburgs Universitair Centrum}\\
 {\it 3590
Diepenbeek, Belgium}}
\maketitle

\ \\
\ \\
\ \\
\begin{abstract}
We show that the dissipative Abelian sandpile on a 
graph $\cal {L}$ can be related
to a random walk on a graph which
consists of $\cal {L}$ extended with a trapping site.
From this relation it can be shown, using exact
results and a scaling assumption, that the dissipative
sandpiles' correlation
length exponent 
$\nu$ always
equals $1/d_{w}$, where $d_{w}$ is the fractal
dimension of the random walker. This leads
to a new understanding of the known result
that $\nu=1/2$ on any Euclidean lattice. Our result
is however more general and as an example we also present
exact data for finite Sierpinski gaskets which
fully confirm our predictions.
\end{abstract}
\newpage
Self organised criticality (SOC) \cite{Bak1,Bak2} is the
phenomenon in which a slowly driven system with many interacting degrees of
freedom evolves spontaneously into a critical state,
characterised by
long range correlations in space and time
(for introductory reviews, see \cite{Jensenbook,Bakbook}).
This phenomenon has by now been recognised (or
conjectured to exist) in
many (models of) natural phenomena such as earthquakes \cite{OFC},
forest fires \cite{FF}, speciation of life \cite{BS},\ldots
Moreover, the possible presence of SOC can be investigated 
in experimentally controlable phenomena such as the
Barkhausen effect \cite{DZ}, rice piles \cite{Gran},
and so on.

A question which is however still poorly understood is what
precisely are the necessary ingredients a system (or
a model) must have for it to become self-organised critical.
For example, one may ask whether or not dissipation of
`energy' (or a similar quantity) destroys long range
correlations. This problem has been studied extensively
in the Olami-Feder-Christensen (OFC) model of
earthquakes \cite{OFC}. Numerical studies originally
seemed to show convincingly that even in the presence of a
small amount of dissipation the model remains critical \cite{OFC-sim}.
Later, it was shown exactly that at least in mean field, the OFC
model is only critical when its energy is conserved \cite{OFC-mf}.
Most recently it was argued on the basis
of a study of branching rates, that the same is true
on a finite dimensional lattice \cite{OFC-br}.

The situation is less controversial for sandpile
models which form {\it the} paradigmatic examples
of systems showing SOC.
In the past decade much progress has been made in the
theoretical understanding of these type of
models. This is especially true for the
Abelian model \cite{Bak1,Bak2},
where, following the original work of Dhar \cite{Dhar},
a mathematical formalism was developped [10--14]
that allows an exact calculation of several properties
of the model such as height probabilities \cite{DharM, Prh},
the upper critical dimension \cite{DcUp} and so on.

The role of conservation (of sand) in sandpile models was first
studied numerically by Ghaffari et al. \cite{GLJ}, who found that
any amount of dissipation destroys the presence of SOC
in the model. Recently, it was proven that indeed on any hypercubic
lattice a non-conservative Abelian sandpile model is not 
critical \cite{Kat1}. 
There exist therefore a correlation length $\xi$ in the system
which diverges when the
dissipation rate goes to zero. This allows the introduction
of an exponent $\nu$ which describes this divergence. Numerically
\cite{GLJ} it was found that $\nu \approx 1/2$ in $d=2$. The same
authors argued on the basis of a renormalisation group calculation
that $\nu=1/2$ on any Euclidean lattice. This result was recently
proven exactly \cite{Kat2}.

In this paper we study further the Abelian sandpile model with
dissipation. We begin by showing that this problem can be related to
that of a suitably defined random walker on a lattice with a trap.
This result is quite general and extends an earlier mapping
between conservative sandpiles and resistor networks (or equivalently
the $q \to 0$ Potts model) \cite{DharM}. Indeed, we will show
that sending the dissipation to zero is equivalent to taking
the long time limit of the random walk problem. It therefore
comes as no surprise that the (sandpile) correlation length exponent
$\nu$ can be related to the exponent $1/d_{w}$ which describes
the asymptotic behaviour of the random walker. Our result 
implies that $\nu=1/2$
for any Euclidean lattice. 
We thus recover in this situation the conclusion of \cite{GLJ, Kat2} but
add a new understanding of it. However our prediction 
is more general and holds also
on, for example, fractal lattices, or for certain types of random sandpiles. 
As an example, we performed 
calculations on
the Sierpinski gasket for which $d_{w}$ is known exactly
($d_{w}=\log{5}/\log{2}$). Our data are consistent
with the prediction $\nu=1/d_{w}$.

The Abelian sandpile model on an arbitrary graph $\cal{L}$ (with
$N$ vertices) is defined
as follows. On each vertex $i$ of the graph, there exists a height
variable $z_{i}$ which assumes integer values and has the
interpretation of energy or number of sand grains at site $i$. The dynamics
of the model consists of two steps. First, we choose any
site $i$ at random, and add one grain of sand to that site,
i.e. $z_{i} \to z_{i}+1$. When at a given site $i$, the
height of sand exceeds a threshold $z_{ic}$, i.e. for $z_{i} > z_{ic}$,
we say that that site becomes unstable and then
the grains of sand on $i$ are distributed among neighbouring sites. This
process, called {\it toppling}, is specified by a matrix $\Delta_{ij}$
such that
\begin{eqnarray}
z_{j} \ \to \ z_{j}-\Delta_{ij}
\end{eqnarray}
The elements $\Delta_{ij}$ satisfy $\Delta_{ii} > 0$ and $\Delta_{ij} 
\leq 0$ when $i \neq j$, and the condition $\sum_{j} \Delta_{ij} \geq 
0$ which garantees that no sand is created in the toppling process.
We will also limit ourselves to cases in which $\Delta$ is symmetric.
Through toppling, neighbouring sites can become unstable and in this
way an {\it avalanche} of topplings is generated. 
A new grain of sand is added only when the avalanche is over,
i.e. all sites are stable again.
Finally, grains of sand can leave the system through certain boundary
sites. These are necessary for the model to reach a steady state
asymptotically.
The number of topplings in a given avalanche $s$ is a random variable
whose distribution $P(s,N)$ is now known to have rich, multifractal
properties \cite{Mario}. In the present paper we will however only 
be interested in the first moment $\langle s \rangle$ of this 
distribution. An exact
expression for this quantity can be obtained as follows \cite{Dhar}.
One first introduces the matrix $G$, which is the inverse of
$\Delta$. The element $G_{ij}$ can be interpreted as the
expected number of topplings at site $j$ when a grain of sand
has been dropped at site $i$ \cite{Dhar}. From this interpretation
one obtains
\begin{eqnarray}
\langle s \rangle = \frac{1}{N} \sum_{i \in {\cal{L}}} \sum_{j \in {\cal{L}}} G_{ij}
\label{2}
\end{eqnarray}
One of the results of this paper will be a scaling expression for 
$\langle s \rangle$ for the dissipative sandpile model.

To continue we will reason further with the case of a graph where each
site (apart from the boundary sites) is connected to a fixed number of 
neighbours $z$. We take $z_{ic}=z, \forall i$. For example, on the
square lattice or on the Sierpinski gasket $z=4$. In the case of
the conservative sandpile model we choose the matrix $\Delta$
as
\begin{eqnarray}
\Delta^{c}_{i,j}=\left\{ \begin{array}{ll}
z&\mbox{if} \ i=j\\
-1&\mbox{i and j are neighbours}\\
0&\mbox{otherwise}
\end{array}\right.
\label{3}
\end{eqnarray}
(in the rest of this paper a superscript c (resp. d) will refer to the
conservative (resp. dissipative) case). On the other hand, following Tsuchiya 
and Katori \cite{Kat1}
, we choose in the case of a dissipative sandpile model
\begin{eqnarray}
\Delta^{d}_{i,j}=\left\{ \begin{array}{ll}
z\gamma\zeta&\mbox{if} \ i=j\\
-\zeta&\mbox{i and j are neighbours}\\
0&\mbox{otherwise}
\end{array}\right.
\label{4}
\end{eqnarray}
with $\gamma>1$. In this way, at each toppling $\zeta z(\gamma-1)$
grains of sand disappear.

With each toppling matrix $\Delta_{ij}$ we can associate a random 
walk problem. To do this we have to extend the graph $\cal{L}$ with one extra
site, denoted as $T$ which, as we will see immediately, will get
the properties of a trap for the random walker. Let ${\cal{L^{\star}}}=
{\cal{L}} \cup T$. We then define a continuous time random walker
on $\cal{L^{\star}}$ using the matrix elements $\Delta_{ij}$ as 
transition rates. More concrete, the rate by which the walker
jumps from $j$ to $i$ is given by $-\Delta_{ij}$
for any two sites on $\cal{L}$. Secondly, the walker jumps from a site
$j \in \cal{L}$ to the trap $T$ with rate 
$\phi_{j}=\sum_{i}\Delta_{ij}$.
Once the walker reaches $T$, it stays there forever.

Let $P(i,k,t)$ be the conditional probability that the walker is
at site $i \in \cal{L}^{\star}$ at time $t$ given that he was in $k$ 
at $t=0$. This conditional probability then evolves according
to the master equation
\begin{eqnarray}
\dot{P}(i,k,t) = - \sum_{j} D_{ij}P(j,k,t)
\label{5}
\end{eqnarray}
where
\begin{eqnarray}
D_{ij} = \left\{ \begin{array}{cl}
\Delta_{ij} & i \in {\cal{L}},\ j \in {\cal{L}},\ \ i \neq j \\
-\phi_{j} & i=T,\ j \in {\cal{L}} \\
0 &\ \mbox{if}\ j=T
\end{array}\right.
\label{7}
\end{eqnarray}
For the diagonal elements we take
\begin{eqnarray*}
D_{jj}= - \sum_{i \in {\cal {L}}^{\star}, i \neq j} D_{ij} =
- \sum_{i \in {\cal {L}}, i \neq j} \Delta_{ij} + \phi_{j} = 
\Delta_{jj}
\end{eqnarray*}
In this way, and because of the conditions we put on the matrix 
$\Delta$,
 $D$ has all the necessary properties (see e.g. \cite{VK}) of a stochastic
 matrix.

To solve the master equation (\ref{5}), it is common (\cite{VK})  to introduce
the Green function $G_{ik}(s)$ which is the Laplace transform of 
$P(i,k,t)$
\begin{eqnarray}
G_{ik}(s) = \int_{0}^{\infty} P(i,k,t) e^{-st}dt
\label{6}
\end{eqnarray}
The Green function obeys the linear equation
\begin{eqnarray*}
\delta_{ik}-sG_{ik}(s) = \sum_{j} D_{ij}G_{jk}(s)
\end{eqnarray*}
whose formal solution is
\begin{eqnarray}
G_{ik}(s)=\left(\frac{1}{s.1+D}\right)_{ik}
\end{eqnarray}
This solution can be written in terms of the eigenvalues 
$\lambda_{\alpha}$ and associated eigenvectors $u_{\alpha}$
of $D$ as
\begin{eqnarray}
G_{ik}(s) = \sum_{\alpha} \frac{1}{(s+\lambda_{\alpha})} 
(u_{\alpha})_{i}(u_{\alpha})_{k}
\label{8}
\end{eqnarray}
Notice that because of the structure (\ref{7}), the spectrum of the 
matrix consists of the spectrum of $\Delta$ and one zero eigenvalue,
which we will denote as $\lambda_{0}$.
The eigenvector associated with this zero eigenvalue is completely
concentrated on the trap, $(u_{0})_{i}=\delta_{i,T}$. 
On the other hand, for the eigenvectors associated with the other
eigenvalues, we have $(u_{\alpha})_{T}=0, \alpha \neq 0$.
Therefore
(\ref{8}) becomes
\begin{eqnarray}
G_{ik}(s) =\frac{1}{s} \delta_{i,T}\delta_{k,T} + \tilde{G}_{ik}(s)
\end{eqnarray}
where
\begin{eqnarray}
\tilde{G}_{ik}(s) = \sum_{\alpha\neq 0}  \frac{1}{(s+\lambda_{\alpha})} 
(u_{\alpha})_{i}(u_{\alpha})_{k}
\label{12}
\end{eqnarray}

We can therefore relate the matrix elements $G_{ij}$ appearing in
(\ref{2}) to the elements of the Green function as
\begin{eqnarray}
G_{ij}=\tilde{G}_{ij}(s=0)
\label{11}
\end{eqnarray}

We now have all the necessary ingredients to discuss the dissipative
sandpiles whose toppling matrix is given in (\ref{4}).
Comparing (\ref{3}) with (\ref{4}) , we immediately see that
\begin{eqnarray}
\Delta^{d}_{ij}= \zeta \Delta^{c}_{ij} + z \zeta (\gamma-1) \delta_{ij}
\label{122}
\end{eqnarray}
This simple relation implies that the eigenvectors of $\Delta^{d}$
and $\Delta^{c}$ are the same and that their eigenvalues are trivially
related by a multiplicative and an additive constant. Therefore
we get for the inverse of $\Delta_{ij}^{d}$ using (\ref{11}) and
the general definition (\ref{12})
\begin{eqnarray}
G^{d}_{ij} = \tilde{G}^{d}_{ij}(s=0) = 
\frac{1}{\zeta}\tilde{G}^{c}_{ij}(s=z(\gamma-1))
\label{14}
\end{eqnarray}
This is our main result. It shows the relation between the dissipative 
sandpile model and the random walker associated
with the {\it conservative} sandpile model but at $s=z(\gamma-1)$. Taking the
conservative limit
$\gamma \to 1$ then corresponds precisely to taking the limit $t \to 
\infty$ in the random walk problem. Since that asymptotic limit
is determined by the scaling exponent $d_{w}$ of the random walk,
it can already be expected that the correlation length exponent
$\nu$ is related to $d_{w}$.

The precise connection between the two exponents can be obtained
as follows. We will calculate $\langle s \rangle$ as given in
(\ref{2}) for the dissipative sandpile defined in (\ref{4}). 
Summing over $i$, using (\ref{14}) and (\ref{6}) we get
\begin{eqnarray}
\sum_{i \in {\cal{L}}} G^{d}_{ik} = \frac{1}{\zeta} \int_{0}^{\infty}
P^{c}_{0}(k,t) e^{-z(\gamma-1)t}dt
\label{15}
\end{eqnarray}
where $P^{c}_{0}(k,t)=\sum_{i  \in {\cal{L}}} P(i,k,t)$ is the
probability that the walker that started at $k$ has not yet been trapped
 at time $t$.
To calculate $P^{c}_{0}(k,t)$ it is important to first consider
the type of random walk that we have to investigate. Because of
the relation (\ref{14}), we have to work with the random walker
associated with the matrix $\Delta^{c}$ of the conservative sandpile.
The resulting random walker is therefore such that only boundary
sites, which in the conservative case are the
only ones where sand leaves the system, are connected with the trap.
Random walks of this type, at least on an Euclidean lattice
are easy to study. We will come back to that below.

First, we consider however the limit $N \to \infty$. In that limit
a random walker starting on a typical site will not be trapped for
any finite $t$, since only the boundary sites at infinity
are connected with the trap. Therefore, one has $P^{c}_{0}(k,t)=1$. Then the
integral in (\ref{15}) can immediately be performed and since
the result does not depend on $k$ we get
\begin{eqnarray}
\langle s \rangle = \frac{1}{\zeta z (\gamma-1)}
\label{16}
\end{eqnarray}
This relation was first derived on the $d=2$ square lattice in
\cite{Kat1} but we now see that it is quite general. In fact,
it shows that the dissipative sandpile is {\it never critical}.

We now turn back to the case that the number of sites on the graph
is finite, in which case $P_{0}^{c}(k,t)$ will be a decreasing
function of time. Explicit results for this quantity can be
obtained on Euclidean lattices with elementary Fourier techniques.
In one dimension one obtains for example
\begin{eqnarray}
P^{c}_{0}(k,t) = \frac{2}{L+1} \sum_{j=1}^{L} \sum_{n=1}^{L} 
e^{-\omega_{n}t} \sin{\left(\frac{n\pi k}{L+1}\right)}
\sin{\left(\frac{n\pi j}{L+1}\right)}
\end{eqnarray}
where $\omega_{n}=2\left[1-\cos(\frac{n\pi}{L+1})\right]$.
Similar results can easily be obtained in higher dimensions.
In the saling limit, $L \to \infty, t \to \infty$, it is easy to
see that this probability is of the form
\begin{eqnarray}
P^{c}_{0}(k,t) \approx \frac{1}{L} F(\frac{k}{L},\frac{t}{L^{2}})
\label{achttien}
\end{eqnarray}
On any Euclidean lattice, the fractal dimension $d_{w}$ of the type of random
walker that we consider here equals $2$. On the basis of 
exact results such as (\ref{achttien}), and on general physical intuition
it can be expected that for the random walkers which are connected
with the trap only through some boundary sites, $P^{c}_{0}(k,t)$
has in general the following scaling behaviour
\begin{eqnarray}
P^{c}_{0}(k,t) \approx \frac{1}{L^{d_{f}}} H(\frac{k}{L},\frac{t}{L^{d_{w}}})
\label{17}
\end{eqnarray}
with $H$ some scaling function.
Here $d_{f}$ is the (fractal) dimension of the graph.

After inserting (\ref{17}) in (\ref{15}), making a suitable change
of variables, and  also performing the resulting sum over $k$ we
finally get the following scaling form for $\langle s \rangle$
\begin{eqnarray}
\langle s \rangle \sim L^{d_{w}} R(L (\gamma-1)^{1/d_{w}})
\label{18}
\end{eqnarray}
where $R$ is a scaling function.

From (\ref{18}), we see that for the conservative case, $\gamma=1$, and on an Euclidean
lattice $\langle s \rangle \sim L^{2}$ which is an old result obtained
by Dhar \cite{Dhar}.
For $\gamma>1$, we conclude from (\ref{18}) that the
exponent $\nu$ that the describes the crossover between dissipative
and conservative sandpiles equals $1/d_{w}$. 

On
an Euclidean lattice, we recover in this way a result 
 first determined with an approximate renormalisation
technique \cite{GLJ} and recently obtained exactly \cite{Kat2}.
Our result (\ref{18}) is however much more general.
The relation $\nu=1/d_{w}$ should also hold on, e.g. fractal lattices.
As an example, we checked
this scaling on finite Sierpinski gasket. In particular, we considered
Sierpinski gaskets of $n$ generations with $n \leq 7$. For each of
these we calculated the matrix $G^{d}_{ij}$ by calculating the
inverse of $\Delta^{d}_{ij}$ using a computer. This we did for
several values of $\gamma$. From $G^{d}$ we then calculated
$\langle s \rangle$ using (\ref{2}). Finally, we plotted 
$\langle s \rangle L^{-d_{w}}$ versus $L (\gamma-1)^{1/d_{w}}$.
These numerically exact results are shown in figure 1.
For the Sierpinski gasket $d_{w}=\log{5}/\log{2}$. As can
be seen the agreement with the scaling prediction is excellent.
Together with the known result for the square lattice case, these
data give very strong support for the conjecture that
$\nu=1/d_{w}$.

\begin{figure}
  \centerline{ \epsfig{figure=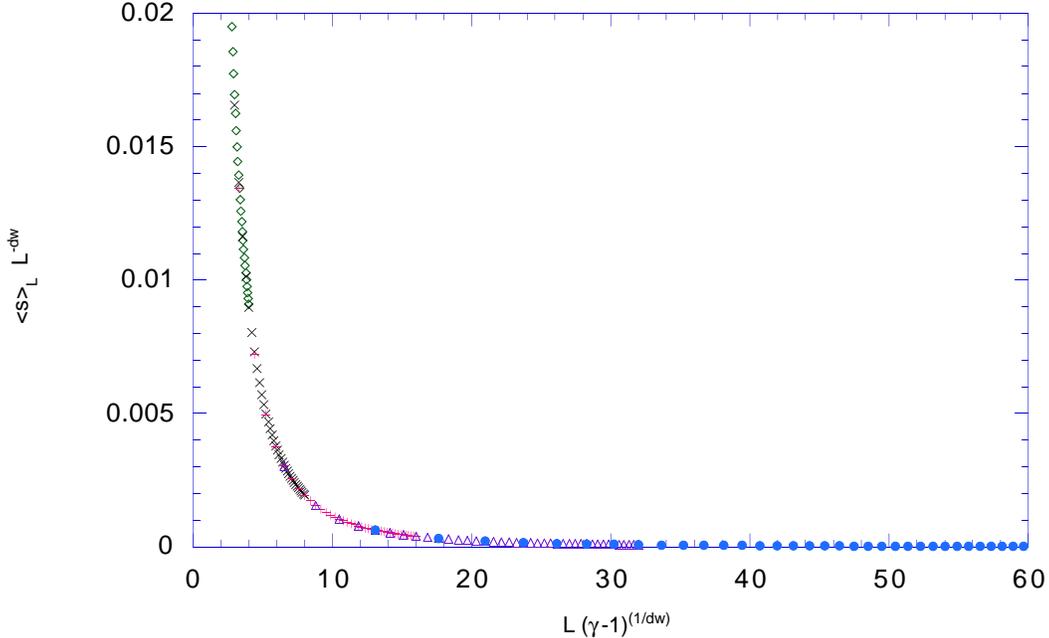,width=14cm}}
\caption{Plot of $\langle s \rangle L^{-d_{w}}$ versus
$L (\gamma-1)^{1/d_{w}}$ for the non conservative
sandpile model on the Sierpinski gasket with $L=2,4,\ldots,128$,
and different values of $\gamma$.}

\label{figure 1}
\end{figure}

One may now ask how general our prediction for $\nu$ is, and whether one
can imagine situations in which it does not hold.
Besides the scaling assumption (\ref{17}), the crucial step
in the derivation is the relation (\ref{122}) between the
conservative and dissipative toppling matrices. 
In a more general case one may consider a toppling matrix
of the form
\begin{eqnarray}
\Delta^{d}_{i,j}=\left\{ \begin{array}{ll}
z\gamma_{i}\zeta&\mbox{if} \ i=j\\
-\zeta&\mbox{i and j are neighbours}\\
0&\mbox{otherwise}
\end{array}\right.
\label{25}
\end{eqnarray}
where $\gamma_{i}$ is site dependent (the same would hold
for a graph in which the coordination number is site
dependent). Let $\gamma_{m}=\min_{i}\gamma_{i}$.
For this case
(\ref{14}) gets replaced by
\begin{eqnarray*}
G^{d}_{ij} = \frac{1}{\zeta} \tilde{G}^{r}_{ij}(s=z(\gamma_{m}-1))
\end{eqnarray*}
where $\tilde{G}^{r}$ is the Green function of a random walker
where from site $i$ one enters the trap with rate
$\zeta z (\gamma_{i}-\gamma_{m})$. It then depends on the
particular values of $\gamma_{i}$ what is the precise
value of $d_{w}$ or even whether $d_{w}$ can still be
defined in a meaningful way. It is also not clear
in which cases a scaling assumption such as (\ref{17})
remains valid. But we expect that for many cases that
are of interest from the sandpile point of view,
our result will hold, eventually with an exponent
$\nu$ that depends on the particular distribution
of $\gamma_{i}$ values.

As an example, take a matrix $\Delta^{d}_{ij}$ whose diagonal
elements are constructed in the following random way.
With probability $p$ we take $\Delta^{d}_{ii}=z\zeta
\gamma_{1}$, while with probability $1-p$ we
take $\Delta^{d}_{ii}=z\zeta\gamma_{2}$. 
In the related random walker, and
for $\gamma_{2}>\gamma_{1}$,
the sites that are connected with the trap are are randomly
distributed over the lattice
with probability $p$. For such a random walk, it has been
shown that in $d=1$, $d_{w}=3$ \cite{WH}. Hence, we
expect that for this kind of dissipative sandpile $\nu=1/3$
(instead of $1/2$) in $d=1$.
We are currently verifying this prediction. The results
will be published elsewhere.

In summary, we have shown how to relate a dissipative
sandpile model with an associated random walker. 
Using results from the theory of random walks we were
then able to show that $\nu=1/d_{w}$ for a large
class of non-conservative sandpile models. This
result is in agreement with the available evidence
on Euclidean lattices and on the Sierpinski gasket.
Since the knowledge on random walks is quite extensive,
we suspect that many interesting phenomena in dissipative
sandpiles can now be obtained using the link with
random walks.

{\bf Acknowledgement} We would like to thank M. Katori
for bringing our attention to this problem and for
useful discussion. We thank J. Hooyberghs for useful
information on random walks with traps. We also
thank V. B. Priezzhev for a critical reading of an earlier
version of this manuscript. Finally, we thank the 
Inter University Attraction Poles for financial support.

\end {document}